\documentclass{article}

\usepackage{PRIMEarxiv}

\usepackage[utf8]{inputenc}
\usepackage[T1]{fontenc} 
\usepackage{hyperref}  
\usepackage{url}    
\usepackage{booktabs}     
\usepackage{amsfonts}    
\usepackage{amsmath}     
\usepackage{nicefrac}   
\usepackage{microtype} 
\usepackage{lipsum}
\usepackage{fancyhdr}     
\usepackage{graphicx}      
\usepackage{float}
\graphicspath{{media/}}    
\usepackage{subcaption}
\usepackage{comment}
\usepackage{lineno}

\usepackage{xcolor}

\usepackage[most]{tcolorbox}

\newtcolorbox{promptbox}[2][]{
  breakable,
  colback=gray!5!white,
  colframe=gray!75!black,
  fonttitle=\bfseries,
  title={#2},
  before upper=\fontsize{10}{12}\selectfont,
  #1
}

\title{Addressing Climate Action Misperceptions with Generative AI}

\author{
  Miriam Remshard\thanks{These authors contributed equally.}, Yara Kyrychenko\footnotemark[1], Sander van der Linden \\
  University of Cambridge \\
  Cambridge, UK\\
  \texttt{\{mr965, yk408, sv395\}@cam.ac.uk} \\
  \And
  Matthew H. Goldberg, Anthony Leiserowitz\\
  Yale University\\
  New Haven, CT, USA\\
  \texttt{\{matthew.goldberg, anthony.leiserowitz\}@yale.edu} \\
  \AND
  Elena Savoia\thanks{These authors share senior authorship.} \\
  Harvard T.H. Chan School of Public Health \\
  Boston, MA, USA \\
  \texttt{esavoia@hsph.harvard.edu} \\
  \And
  Jon Roozenbeek\footnotemark[2]\\
  University of Cambridge \\
  Cambridge, UK\\
  \texttt{jjr51@cam.ac.uk} \\
}

\begin{document}
\maketitle

\vspace{-1cm}
\begin{abstract}
Mitigating climate change requires behaviour change. However, even climate-concerned individuals often hold misperceptions about which actions most reduce carbon emissions. We recruited 1201 climate-concerned individuals to examine whether discussing climate actions with a large language model (LLM) equipped with climate knowledge and prompted to provide personalised responses would foster more accurate perceptions of the impacts of climate actions and increase willingness to adopt feasible, high-impact behaviours. We compared this to having participants run a web search, have a conversation with an unspecialised LLM, and no intervention. The personalised climate LLM was the only condition that led to increased knowledge about the impacts of climate actions and greater intentions to adopt impactful behaviours. While the personalised climate LLM did not outperform a web search in improving understanding of climate action impacts, the ability of LLMs to deliver personalised, actionable guidance may make them more effective at motivating impactful pro-climate behaviour change.
\end{abstract}

\vspace{0.5cm}

 Climate change poses significant risks to the environment and its inhabitants \cite{lee_ipcc_nodate}. As extreme weather events have become more frequent \cite{myhre_frequency_2019}, and media coverage of the issue has increased in countries such as the US and the UK \cite{hase_climate_2021}, public awareness and concern about climate change have grown. For example, the proportion of the US public classified as ``Alarmed'' about climate change increased by 11 percentage points between 2014 and 2024. This shift in public perception could lead to positive developments in addressing the issue, as ``Alarmed'' individuals are more likely to take and support pro-climate actions and policies \cite{leiserowitz_global_nodate}. However, despite this growing concern, US commercial and residential greenhouse gas emissions have remained relatively stable over the past 30 years \cite{environmental_protection_agency_commercial_2025}. Numerous barriers may prevent even motivated people from taking effective pro-climate actions \cite{gifford_dragons_2011, latkin_perceived_2023}. For example, climate-concerned individuals have been shown to misjudge the impacts of pro-climate behaviours, systematically overestimating low-impact actions and underestimating high-impact behaviours \cite{remshard_understanding_2025}. 

Considering the multiple psychological factors required to motivate action on climate change \cite{stern_value-belief-norm_1999}, and the fact that a large percentage of the public is already concerned about the issue \cite{leiserowitz_international_nodate}, a key bottleneck appears to be how concern translates into effective behaviour. Climate-concerned individuals tend to engage more in actions that they perceive as effective at reducing CO$_2$ emissions \cite{remshard_understanding_2025}, suggesting that correcting misjudgements of the impacts of climate actions could lead to higher impact behaviours. Recent evidence supports this view: providing accurate information about the impacts of pro-climate behaviours was shown to improve impact estimates and increase commitment to take high-impact actions \cite{goldwert_climate_2025}. Yet although broad patterns in the relative effectiveness of climate actions are well established, the specific behaviours that matter most vary among individuals. For example, purchasing an electric vehicle would not constitute a high-impact action for someone without a car, as the carbon reduction potential of living car-free is greater than that of owning an electric vehicle \cite{ivanova_quantifying_2020}. More generally, the volume and heterogeneity of everyday behaviours that shape individual carbon footprints limit the usefulness of generic educational interventions.

With the advent of generative artificial intelligence (AI), this challenge can be addressed with tailored interventions. Personalised large language models (LLMs) can utilise user characteristics to tailor information to specific users' needs \cite{chen_when_2023}. Indeed, a multi-study exploration of the persuasive effects of personalised ChatGPT messages found that personalisation based on psychological variables influenced participants' preferences for advertisements and climate change appeals, as well as their willingness to pay for a consumer product \cite{matz_potential_2024}. Although research on political persuasion has been more sceptical of the persuasive capacities of personalised messaging through generative AI \cite{argyle_testing_2025, hackenburg_levers_nodate}, it has been argued that in the realm of climate misperceptions, personalised persuasion has the potential to be particularly powerful because ``[p]ersonalised recommendations can empower consumers to adopt low-carbon technologies by suggesting options that align with their needs while minimising their environmental impact'' (p. 3, \cite{stern_green_2025}). In other words, a conversation with an LLM about climate actions might promote more accurate perceptions of the true emissions savings associated with different pro-environmental behaviours and encourage higher impact behaviours, as it can highlight high-impact actions that are feasible and relevant to the individual.

Conversing with personalised LLMs has a further advantage over traditional educational approaches: Critical perspectives on intervention research have raised the concern that intervention attractiveness and uptake potential are rarely considered when assessing intervention efficacy\cite{roozenbeek_beyond_2024}. Consequently, if people are unlikely to engage with an intervention outside of a paid study, then the intervention's actual effectiveness will be far lower than the efficacy that has been theoretically established. However, generative AI is gaining user appeal, with the share of US adults who have used ChatGPT doubling from 18\% in 2023 to 34\% in 2025, and rising to 58\% among adults under 30 \cite{sidoti_34_2025}. It is thus plausible that climate-concerned individuals will increasingly ask LLMs about the relative environmental impacts of different actions they could take.

Given this potential for generative AI as a tool for pro-environmental communication, several studies have explored its persuasive capacities for reducing climate scepticism. Czarnek et al. \cite{czarnek_addressing_2025} and Hornsey et al. \cite{hornsey_promise_2025} found that conversations with ChatGPT reduced specific reservations and general climate change scepticism, respectively. However, while the former study reported that approximately 36\% of the effect remained one month post-intervention, the latter reported full decay after two weeks. Generative AI has also been shown to successfully modify the headlines of climate science articles to foster greater content engagement among sceptical audiences \cite{bago_using_2025}.

However, most prior work (e.g., \cite{costello_durably_2024, czarnek_addressing_2025}) employs a rigid AI-driven conversational protocol with pre- and post-intervention measurements to assess belief change. This does not reflect how people typically engage with chatbots, with most conversations being human-initiated, and risks introducing demand characteristics. Therefore, we designed our intervention to be more akin to how a user, especially a climate-conscious one, would engage with an LLM of their own accord. Specifically, participants initiated the conversation with the goal of learning about climate actions and were unaware that they would receive a subsequent knowledge test. 

Moreover, the control conditions used in previous studies ranged from conversations with an LLM on an unrelated topic \cite{costello_durably_2024, czarnek_addressing_2025} to reading a press release synthesising a climate science report \cite{hornsey_promise_2025}. Expanding on this work, we compare conversing with generative AI and searching the web on the same topic. This comparison is particularly informative, as (1) learning from LLMs has been shown to produce shallower knowledge than web searches \cite{melumad_experimental_2025} and (2) the environmental cost of LLMs in terms of electricity use and carbon emissions is greater: a Google search consumes approximately 0.30 Wh, while a short GPT-4o query uses roughly 40\% more (0.42 Wh) \cite{jegham_how_2025}. Therefore, our study aimed to explore whether personalised generative AI messages outperform the more environmentally friendly alternative of a web search. 

\begin{figure}[t!]
    \centering
    \includegraphics[width=\linewidth,trim={0 0 0 0}]{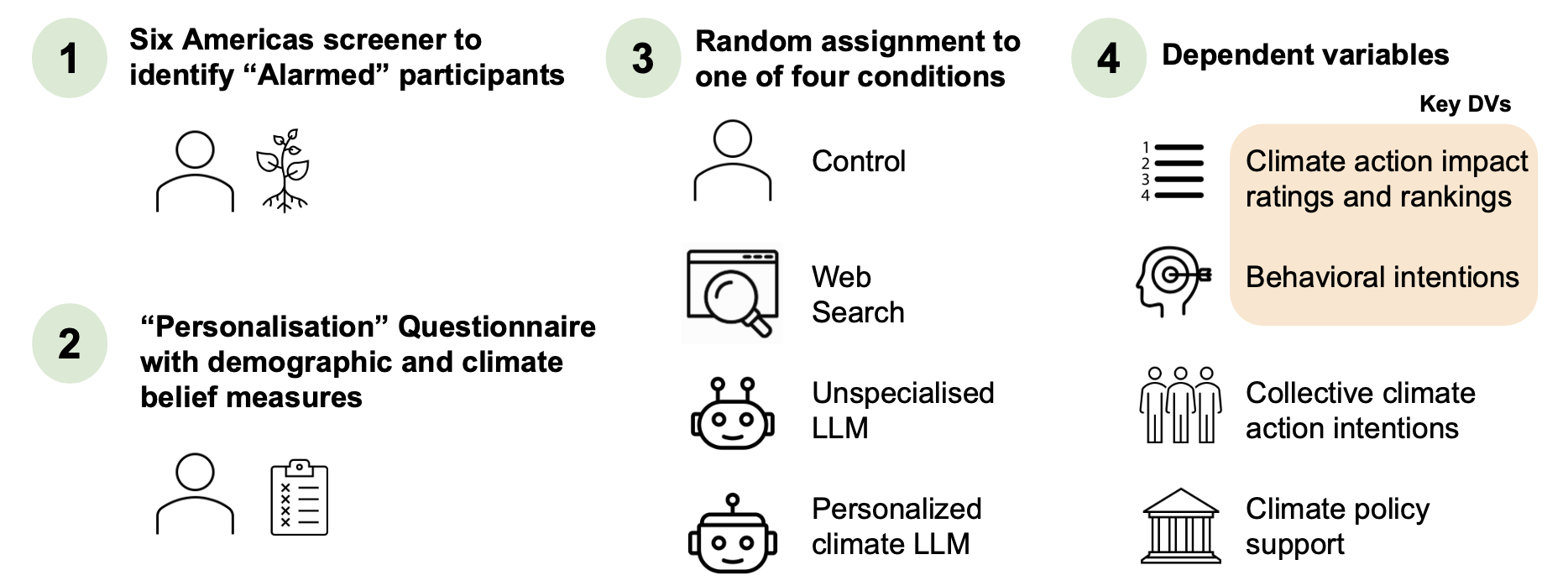}
    \caption{Study procedure as outlined in detail in the Methods section.}
\end{figure}

Consequently, the current study aimed to understand the potential of generative AI in promoting intentions to engage in high-impact pro-climate behaviours by exploring the capacity of LLM conversations to address climate action misperceptions among \textit{climate-concerned} audiences. We compared the effect of LLM conversations against searching the web for relevant information. Overall, in a study with 1201 climate-concerned individuals, we found that interacting with a personalised climate LLM reduces climate action misperceptions and increases intentions to engage in feasible, high-impact behaviours. 

\section*{Results}
To investigate whether generative AI can improve impact assessments and behavioural intentions among climate-concerned individuals, we recruited 1201 participants concerned about climate change from Prolific.\footnote{For 94 participants, we were unable to match their survey data to their interaction with the intervention web app. We were therefore unable to examine their interactions with the chatbot or their web searches and could not know for sure that they participated in the intervention as intended. To ensure that this did not adversely affect the efficacy of the different experimental conditions, we ran our analyses with only the 1107 participants with matched data entries. However, the results were not meaningfully different for any of our key dependent variables (see Supplementary Information \textit{10}), so we report the analyses with all 1201 participants below.} Participants were assigned to one of four conditions (see Methods for further details): 

\begin{itemize}
    \item[\textit{pLLM}] \textbf{Personalised climate LLM}: Participants discussed the impacts of climate actions with an LLM equipped with climate knowledge and tailored to provide personalised responses.
    \item[\textit{uLLM}] \textbf{Unspecialised LLM}: Participants discussed the impacts of climate actions with an unspecialised LLM.
    \item[\textit{WS}] \textbf{Web search}: Participants ran a web search to explore the impacts of climate actions. 
    \item[\textit{TC}] \textbf{True control}: Participants in this condition completed all study measures without any intervention.  
\end{itemize}

Subsequently, all participants indicated their intentions to take 14 pro-environmental behaviours and three collective climate actions going forward, rated the same 14 behaviours as having a small, moderate, or large effect on emission reductions and ranked their impacts in terms of reducing personal greenhouse gas emissions, reported their support for nine climate policies, indicated their use and knowledge of AI, and completed AI feedback measures.  

We pre-registered the following hypotheses (AsPredicted, \#225994):
\begin{itemize}
    \item[$H_{1}$:] pLLM will promote more accurate impact estimates of climate actions \footnote{The accuracy of participants' impact assessments of climate actions was measured using both participants' impact rankings and three-point Likert scale ratings, in line with \cite{remshard_understanding_2025}. See Methods for more information.} than WS, uLLM or TC.
    \item[$H_{2}$:] pLLM will promote greater willingness to engage in feasible, high-impact pro-climate behaviours than  WS, uLLM or TC.
\end{itemize}

Additionally, we asked the following research questions:
\begin{itemize}
    \item[$RQ_{1}$:] (If $H_{2}$ is confirmed): Is the effect on behavioural intentions mediated by more accurate climate action impact estimates?
    \item[$RQ_{2}$:] (If $H_{1}$ and $H_{2}$ are confirmed): Are these effects mediated by AI knowledge and frequency of use?
    \item[$RQ_{3}$:] Does the pLLM promote greater support of pro-climate policies compared to the other conditions?
    \item[$RQ_{4}$:] Does the pLLM foster a stronger willingness to engage in collective climate action compared to the other conditions?
\end{itemize}

\begin{figure}[t!]
    \centering

    \vspace{-0.5cm}

    \begin{subfigure}[t]{.5\textwidth}
        \centering
        \caption{\textbf{a} Impact Ranking Accuracy}
        \includegraphics[width=\linewidth,trim={0 0 0 0}, clip]{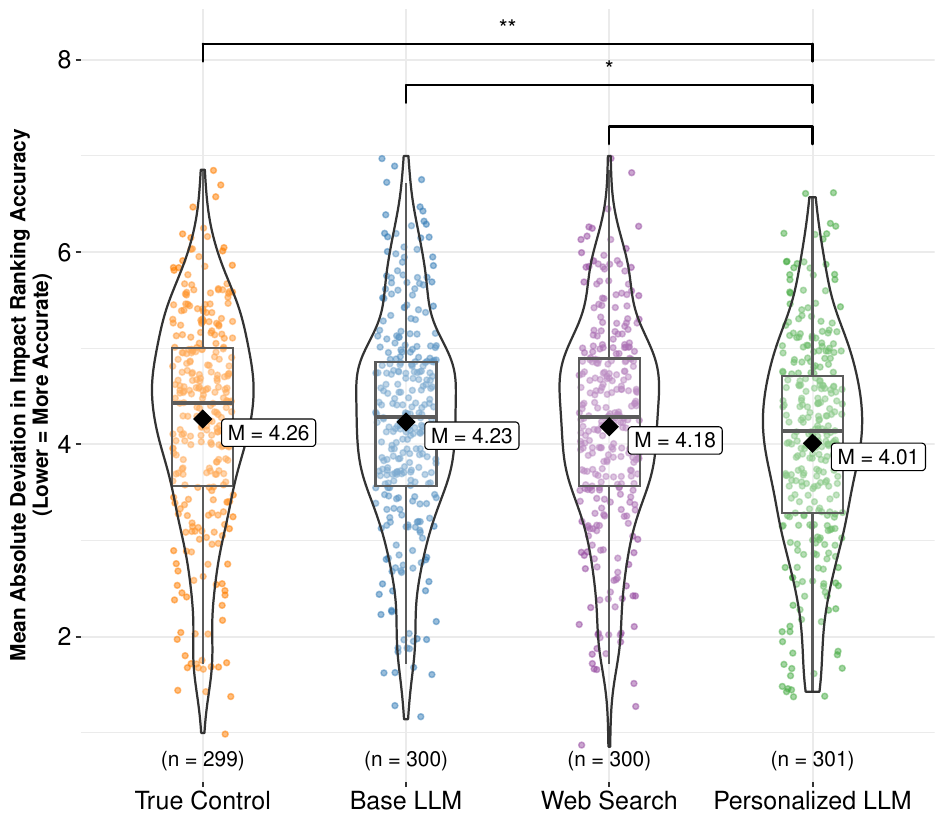}
        \label{fig:impact_rankings}
    \end{subfigure}%
    \begin{subfigure}[t]{.5\textwidth}
        \centering
        \caption{\textbf{b} Behavioural Intentions}
        \includegraphics[width=\linewidth,trim={0cm 0 0 0}, clip]{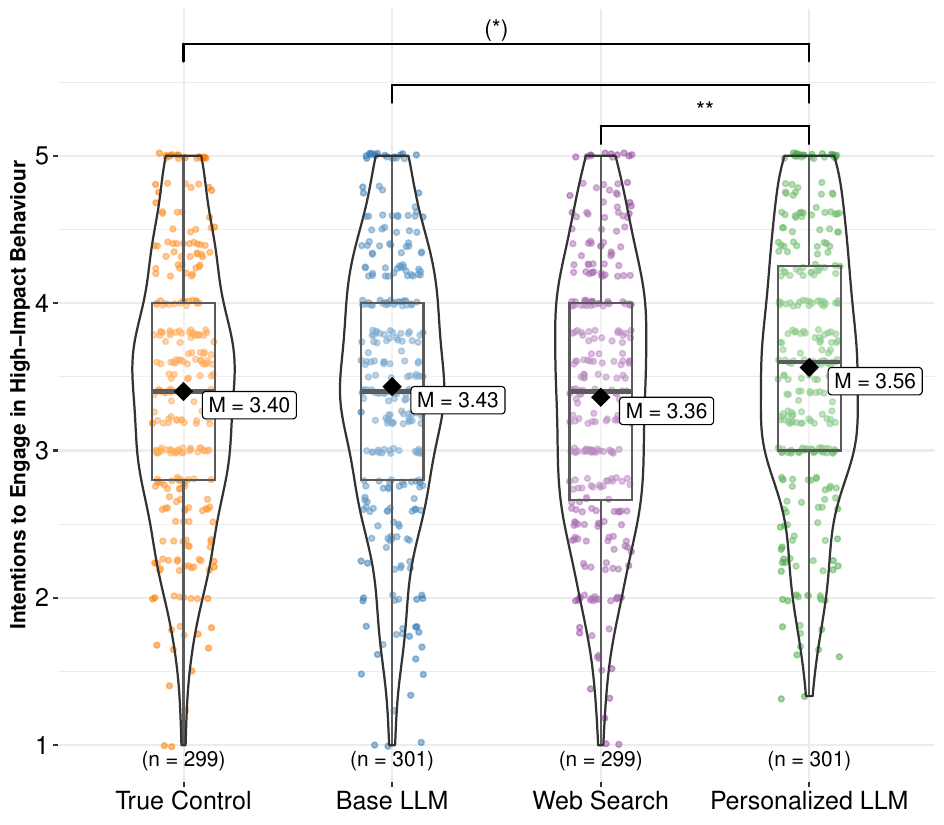}
        \label{fig:beh_intentions}
    \end{subfigure}
    \vspace{-0.5cm}

    \caption{
Comparison across conditions for \textbf{a} impact ranking accuracy and \textbf{b} intentions to engage in high-impact climate actions.
For both panels, jittered points represent individual participants, bold diamond markers indicate condition means (M), and numeric labels report the mean value.
Boxes show the interquartile range, with the central line indicating the median; whiskers extend to 1.5$\times$IQR.
Horizontal brackets denote pairwise comparisons against the Personalized LLM condition.
${}^{*}$ indicates $p_{\text{Holm}} < .05$; ${}^{**}$ indicates $p < .01$ or $p_{\text{Holm}} < .02$.
Asterisks shown in parentheses indicate effects that become marginally non-significant after Holm adjustment.
}

    \label{fig::main}
\end{figure}


We pre-registered specific comparisons reflecting our primary hypothesis using Student's t-tests.\footnote{We compared the key intervention to each control condition individually using pre-specified t-tests rather than an omnibus ANOVA, as our primary estimand concerns intervention–control contrasts and not differences among control conditions. However, running one-way ANOVAs with Tukey's HSD post hoc tests produces highly comparable results (see Supplementary Information \textit{12}). All the assumptions required for t-tests were met (see Supplementary Information \textit{11}).} We adjusted p-values using the Holm correction to reduce the risk of false positive findings. While there were no significant group differences when testing \textit{$H_{1}$} using participants' three-point impact ratings, we did find that pLLM participants' impact rankings were significantly more accurate (\textit{M}$_{\text{pLLM}}$ = 4.01, \textit{SD} = 1.09) with smaller deviations from the true rankings than those in the TC (\textit{M}$_{\text{TC}}$ = 4.26, \textit{SD} = 1.12, \textit{t}(597.22) = 2.78, \textit{p}$_{\text{Holm}}$ = .017, \textit{d} = 0.23) and uLLM conditions (\textit{M}$_{\text{uLLM}}$ = 4.23, \textit{SD} = 1.12, \textit{t}(598.65) = 2.45, \textit{p}$_{\text{Holm}}$ = .029, \textit{d} = 0.20). However, when comparing impact ranking accuracy between the pLLM and WS conditions (\textit{M}$_{\text{WS}}$ = 4.18, \textit{SD} = 1.11), the difference failed to reach statistical significance (\textit{t}(598.86) = 1.91, \textit{p} = .057, \textit{d} = 0.16).

Next, we ran additional Student's t-tests with Holm's p-value correction to examine whether the pLLM also increased intentions to engage in feasible, high-impact pro-climate behaviours compared to the control conditions (\textit{$H_{2}$})\footnote{To compare the effects of the four conditions for only \textit{feasible} climate actions, we excluded responses where participants indicated that a behaviour was ``Not applicable'' to them or that they had already taken (for one-off actions, such as installing heat pumps).}. The pLLM (\textit{M}$_{\text{pLLM\_high}}$ = 3.56, \textit{SD} = 0.89 and \textit{M}$_{\text{pLLM\_moderate}}$ = 3.50, \textit{SD} = 1.06) promoted greater intentions to engage in both high- and moderate-impact actions compared to the WS condition (\textit{M}$_{\text{WS\_high}}$ = 3.36, \textit{SD} = 0.91, \textit{t}(597.71) = -2.75, \textit{p}$_{\text{Holm}}$ = .018, \textit{d} = -0.22 and \textit{M}$_{\text{WS\_moderate}}$ = 3.26, \textit{SD} = 1.15, \textit{t}(592.07) = -2.66, \textit{p}$_{\text{Holm}}$ = .024, \textit{d} = -0.22). Participants in the pLLM condition also reported greater intentions to take high-impact actions than TC participants (\textit{M}$_{\text{TC\_high}}$ = 3.40, \textit{SD} = 0.89, \textit{t}(598) = -2.24, \textit{p} = .025, \textit{d} = -0.18). However, this effect became marginally non-significant after Holm's correction for multiple comparisons (\textit{p}$_{\text{Holm}}$ = .051). There was no significant difference for intentions to engage in high-impact behaviours (\textit{p} = .077) when comparing the pLLM and uLLM conditions (\textit{M}$_{\text{uLLM\_high}}$ = 3.43, \textit{SD} = 0.90). There were also no further treatment effects for moderate- or low-impact actions, which aligned with our hypothesis that the pLLM would foster intentions to engage specifically in \textit{feasible, high-impact} behaviours.

Having established that pLLM participants were better at ranking climate action impacts and reported greater intentions to engage in feasible, high-impact behaviours than TC participants, we examined whether the pLLM's effect on behavioural intentions was mediated by improved impact assessments using structural equation modelling with 5000 bootstrap samples (\textit{$RQ_{1}$}). While the direct effect of condition on behavioural intentions remained significant (\textit{B} = 0.18, SE = 0.07, \textit{p} = .013), the indirect effect via impact ranking accuracy was not significant (\textit{B} = -0.02, SE = 0.01, \textit{p} = .078) indicating that the pLLM effect on behavioural intentions was not mediated by improved climate action impact rankings. 

We also used structural equation modelling with 5000 bootstrap samples to explore whether the pLLM's effects on improved impact rankings and greater intentions to engage in feasible, high-impact actions compared to TC participants were mediated by AI knowledge or frequency of use (\textit{$RQ_{2}$}). However, the indirect effects of condition on impact ranking accuracy via both AI knowledge (\textit{B} = 0.01, SE = 0.01, \textit{p} = .371) and frequency of use (\textit{B} = 0.02, SE = 0.01, \textit{p} = .148), as well as the indirect effects of condition on high-impact behavioural intentions via AI knowledge (\textit{B} = 0.01, SE = 0.01, \textit{p} = .458) and frequency of use (\textit{B} = 0.02, SE = 0.01, \textit{p} = .092) were non-significant, indicating no mediation. 

Finally, we examined whether pLLM conversations affected related constructs, such as support for pro-climate policies (\textit{$RQ_{3}$}) or willingness to engage in collective climate actions (\textit{$RQ_{4}$}). However, there was no significant difference between support for climate policies in the pLLM condition (\textit{M}$_{\text{pLLM}}$ = 85.82, \textit{SD} = 13.82) and the uLLM (\textit{M}$_{\text{uLLM}}$ = 84.60, \textit{SD} = 13.57, \textit{t}(599.81) = -1.09, \textit{p} = .276, \textit{d} = -0.09), WS (\textit{M}$_{\text{WS}}$ = 85.62, \textit{SD} = 13.32, \textit{t}(598.33) = -0.18, \textit{p} = .858, \textit{d} = -0.01) or TC (\textit{M}$_{\text{TC}}$ = 85.30, \textit{SD} = 14.10, \textit{t}(597.57) = -0.46, \textit{p} = .647, \textit{d} = -0.04) conditions. While pLLM participants did display greater willingness to engage in collective actions (\textit{M}$_{\text{pLLM}}$ = 4.06, \textit{SD} = 0.89) than WS participants (\textit{M}$_{\text{WS}}$ = 3.90, \textit{SD} = 0.93, \textit{t}(597.98) = -2.17, \textit{p} = .031, \textit{d} = -0.18), this effect failed to reach significance following Holm's correction (\textit{p}$_{\text{Holm}}$ = .092). There was no significant difference when comparing pLLM participants against those in the uLLM (\textit{M}$_{\text{uLLM}}$ = 3.94, \textit{SD} = 0.98, \textit{t}(594.51) = -1.58, \textit{p} = .115, \textit{d} = -0.13) or TC (\textit{M}$_{\text{TC}}$ = 4.04, \textit{SD} = 0.97, \textit{t}(592.97) = -0.26, \textit{p} = .797, \textit{d} = -0.02) conditions. 

\section*{Discussion}
In a study with 1201 climate-concerned US individuals, we examined whether a conversation with a personalised LLM equipped with climate knowledge corrected climate action misperceptions and promoted intentions to engage in feasible, high-impact behaviours compared to conversing with an unspecialised LLM, a web search, or no intervention. We found partial support for our hypotheses. 

The personalised climate LLM improved people's ability to rank climate actions according to their impact compared with the unspecialised LLM or not receiving any intervention, but not compared with searching the web. The fact that the personalised climate LLM outperformed the unspecialised LLM, but not the web search, suggests that default chatbots might provide less relevant information even to motivated users, and also aligns with research indicating that web searches foster deeper learning by necessitating more active, self-directed participation compared with passively receiving information through generative AI \cite{melumad_experimental_2025}. 

However, in terms of promoting intentions to engage in feasible, high-impact climate actions, the personalised climate LLM outperformed the web search and true control conditions (although the latter effect became marginally non-significant following Holm's correction), but not the unspecialised LLM. The appeal of using generative AI to promote higher impact climate action is argued to lie in its ability to personalise recommendations, helping consumers identify effective options that align with their needs \cite{stern_green_2025}. This is why the main intervention included a personalisation prompt, so that the LLM would only highlight effective actions that are \textit{feasible} for participants. However, even the unspecialised LLM was capable of personalisation, dynamically adjusting behavioural recommendations based on the information provided by the user during the chat. Therefore, our results suggest that LLMs may generally be effective tools for motivating impactful, pro-climate behavioural intentions even when the LLM is not specifically prompted to do so. 

In short, the experimental conditions had varying effects on improving knowledge of action impacts and intentions to engage in high-impact behaviours: Personalised climate LLM conversations and web searches fostered impact knowledge, whereas conversations with generative AI (both specialised and unspecialised) promoted high-impact behavioural intentions. Beyond identifying this distinct pattern of effects for knowledge versus behavioural outcomes, we found that the effect of the personalised climate LLM on behavioural intentions was not mediated by improved impact rankings. This is notable given prior research demonstrating a correlation between behavioural engagement and perceived action impact \cite{remshard_understanding_2025}, suggesting that climate-concerned individuals might tend to prioritise behaviours they view as effective. However, our findings indicate that distinct cognitive processes might be involved in correcting knowledge of action impacts versus fostering impactful behavioural intentions, with the personalised climate LLM being the only condition to influence both. Thus, while AI has been suggested to reduce conspiracy beliefs primarily by providing factual information \cite{costello_just_2025}, the absence of mediation by improved action impact rankings suggests that, when motivating effective pro-climate behavioural intentions, AI has persuasive capacities that do not stem solely from its ability to increase knowledge.

These results have implications for climate change communication and AI: Our findings show that current state-of-the-art AI models do not effectively improve knowledge of action impacts without climate-specific modifications. We created a personalised and climate-knowledge-enhanced LLM via prompting; however, other LLM alignment techniques might be more robust and effective in real-world applications, including aligning LLMs with climate change communication principles via Constitutional AI \cite{bai_constitutional_2022,kyrychenko2024c3ai} and using retrieval-augmented architectures to improve factual accuracy \cite{guu_retrieval_2020}. Additionally, the differential effects of web search and LLMs on knowledge and motivation suggest that AI overviews on search platforms, if designed appropriately, could provide the best of both worlds by combining the knowledge gains of web searches with the motivational capacities of (even unspecialised) LLMs. Finally, our results highlight the importance of targeting behavioural intentions or, ideally, behaviours directly \cite{dablander_expressing_2025} in climate action interventions, as increased knowledge of action impacts does not necessarily translate to effective behavioural intentions.

Nonetheless, our study has several limitations. Notably, the personalised climate LLM was prompted to personalise its responses based on participants' answers to a set of questions and equipped with knowledge of the emissions savings associated with different pro-climate behaviours. By combining personalisation and climate knowledge within the main treatment condition, we cannot determine how much either factor drove the effects. Further research is needed to clarify the contributions of personalisation and climate knowledge in equipping generative AI to address action impact misperceptions and promote effective behavioural intentions. 

Second, the effects on impact assessments were only found using the ranking measure. When comparing participants' accuracy in rating climate actions as having small, moderate, or large effects on greenhouse gas emissions, there was no significant difference between the groups. While this could indicate that the intervention effects on impact assessments were inconsistent, it more likely reflects limited scale sensitivity when presenting only three impact categorisation options compared to 14 potential rankings. This points towards a wider gap in the field; namely, that there is no agreed-upon, psychometrically validated approach to measuring climate action impact misperceptions. The development of such a scale could facilitate more reliable research into strategies for correcting impact misjudgements. 

Finally, while personalised climate LLM conversations were shown to foster greater intentions to engage in high-impact pro-climate behaviour, we cannot establish whether this relates to changes in participants' \emph{actual} behaviour. Although a well-established phenomenon across issue domains \cite{conner_understanding_2022, sheeran_intentionbehavior_2016}, studies have increasingly demonstrated sizeable ``intention-behavior gaps'' \cite{sheeran_intentionbehavior_2002} in the domain of pro-climate behaviours \cite{goldberg_predicting_2025, osberghaus_intention-behavior_2025}. Consequently, there has been a call to move from measuring intentions to measuring actual behaviour, or at least to establish the extent to which the two are related \cite{dablander_expressing_2025}. However, because the behaviours we examined were either rarely actionable (e.g., installing heat pumps) or long-term in scale (e.g., following a vegetarian diet), we resorted to measuring behavioural intentions. 

In sum, our study shows that conversations with a climate-informed LLM that personalises recommendations toward feasible actions can correct misperceptions about the impacts of pro-climate behaviours and promote intentions to engage in more effective actions. This approach outperformed searching the web, which improved impact knowledge but did not shift behavioural intentions, and interacting with an unspecialised LLM, which increased intentions to engage in high-impact actions without improving impact rankings. Together with the fact that improved rankings of action impacts did not mediate behavioural intentions, this pattern indicates that the cognitive processes underlying the acquisition of climate action knowledge and those guiding behavioural prioritisation may be meaningfully distinct.

\section*{Methods}
This study was approved by both the Institutional Review Board of the Harvard T.H. Chan School of Public Health (IRB25-0233) and the King's College London Research Ethics Office (MRA-24/25-46686) and pre-registered on AsPredicted (\#225994). All materials, data, and analysis code needed to reproduce this study are available on our Open Science Framework (OSF) page at \href{https://osf.io/9yda3}{osf.io/9yda3}.

\subsection*{Sample}
To ensure that only participants who were classified as ``Alarmed'' about climate change according to the well-validated Six Americas audience segmentation \cite{leiserowitz_global_2009} were invited to participate in the study, we first conducted an initial screener survey. We used built-in Prolific screener questions to ensure that the screener survey was advertised only to Prolific users who are very concerned about environmental issues and believe in climate change. After providing informed consent, participants then completed a CAPTCHA before indicating to what extent they think climate change is or is not happening on a 10-point Likert scale (1 = ``I strongly believe climate change is NOT happening'' to 10 = ``I strongly believe climate change IS happening'', \textit{M} = 9.36, \textit{SD} = 1.22). They then completed the four-item Six Americas Super Short Survey \cite{chryst_global_2018} before being thanked for their participation and informed that, depending on their responses, they might be invited to participate in the main study. We screened 2725 Prolific users, comprising 1997 Alarmed, 550 Concerned, 147 Cautious, five Disengaged, 11 Doubtful and 15 Dismissive participants. Only those classified as ``Alarmed'' were re-invited to participate in the main study. 

We had pre-registered that we would collect data from 300 participants per condition, for a total of 1200 participants, to allow for detecting effects of Cohen's \textit{d} around 0.40 with a power of at least .99. However, due to accidentally over-recruiting participants for one of the conditions, we decided to match all other conditions, resulting in a total sample size of 1204 participants with 301 participants per condition. Finally, removing the second submissions of three participants who completed the study twice resulted in a final dataset of \textit{N} = 1201 participants with \textit{n} = 301 participants in the both the personalised and unspecialised LLM conditions, \textit{n} = 300 in the web search condition, and \textit{n} = 299 true control participants. Although in line with Prolific guidance, we removed participants who failed both attention checks in our study, 48 of the 1201 participants in our final sample failed one of the two checks.

\subsection*{Experimental Conditions}
After providing informed consent again, completing another CAPTCHA, and agreeing not to use external AI chatbots to complete the study, participants were randomly allocated to one of four conditions. Participants in all conditions first answered the same ``personalisation'' questions, but their responses were only passed to the personalised climate LLM.

Prior work has attempted to design a personalised LLM conversational intervention to increase pro-environmental behaviour, but found that the personalised LLM can backfire, reducing participants' pro-environmental intentions and sustainable choice preferences as compared to a static message control \cite{doudkin_ai_2025}. Therefore, we carefully selected the personalisation questions relevant to the intervention's goals by choosing those that were highly predictive of impact rankings based on previously available survey data \cite{remshard_understanding_2025}.

Participants answered a set of demographic questions relating to their age (\textit{M} = 39.8, \textit{SD} = 12.9), gender, level of education, and political orientation. Participants were also asked to indicate whether they live in an urban, suburban, rural, or other type of community; whether they rent or own their home; whether they live rent-free or have a different home set-up; and whether they own or lease a car, or do neither. They were then invited to describe any actions that they take to address climate change and to indicate which they believe is the single most effective action that they can take to reduce carbon emissions that contribute to climate change in a textbox. Finally, they used a slider bar to show what percentage of plastic produced they believe gets recycled, before indicating their agreement with the following four statements on a five-point Likert scale: (1) ``Electric vehicles don't have enough range to handle daily travel demands'', (2) ``The fossil fuel industry is trying to shift the blame away from themselves by emphasizing the importance of individual climate action'', (3) ``The use of aerosol spray cans is a major cause of climate change'', and (4) ``Lab-grown meat produces up to 25 times more CO2 than real meat.'' For the descriptive statistics of all survey measures, see Supplementary Information \textit{2}.

Subsequently, participants in the personalised climate LLM condition were invited to discuss the impacts of climate actions with an LLM equipped with climate knowledge. Specifically, they were given examples of the types of questions to ask the LLM, such as ``What are the most effective actions to reduce my carbon emissions?'', ``What's better for the environment: a year of vegetarianism or skipping one transatlantic flight?, ``How do the emissions saved by switching to an EV compare to recycling for a year in terms of trees planted?''. They were also instructed that they would have to respond at least five times and at least three minutes had to elapse before they would be able to proceed with the rest of the study. The conversations lasted on average 5.82 minutes (\textit{SD} = 3.59), spread over 11.30 turns (\textit{SD} = 2.89). 

In the personalised climate LLM condition, the LLM would also encourage participants to ask (follow-up) questions about the impacts of climate actions in its responses. Additionally, the LLM was equipped with a list of climate actions \cite{remshard_understanding_2025} and its corresponding impact classifications \cite{ivanova_quantifying_2020}. It was instructed to use participants' responses to the previous questions to personalise responses, encouraging effective and feasible pro-climate behaviours (e.g., avoid promoting installing heat pumps if the participant does not own their home, or avoid discussing EVs with participants who already live car-free).

Participants in the unspecialised LLM condition were given the same instructions for interacting with the LLM and the same examples of questions to ask the chatbot. However, the chatbot they interacted with was just the baseline LLM, not equipped with any additional knowledge beyond a prompt to unify response formatting. For detailed information about the LLM prompts, see Supplementary Information \textit{13}. Both LLM conditions employed GPT-4.1 (gpt-4.1-2025-04-14).

In the web search condition, participants' instructions were also kept as similar as possible, except that they were asked to google information about the impact of climate actions rather than interact with a chatbot. Consequently, the instructions were modified minimally such that they were asked to type questions, such as the example questions listed above, in the search box to search Google, and informed they had to click on at least five links – rather than respond five times – before they would be able to proceed. Because Google Search now includes AI-generated overviews that could confound our results, we developed a custom web application to retrieve search results without the AI overview.

Finally, participants in the true control condition were informed that they had randomly not been selected to have a conversation with the chatbot after completing the initial set of questions. They were then immediately directed to complete the remaining study measures. 

\subsection*{Measures}
Our dependent variables were derived from measures previously used in Remshard et al. \cite{remshard_understanding_2025}. After completing the intervention tasks according to their conditional assignment, all participants were invited to rate and rank the impacts of 14 pro-climate behaviours, and to report the extent to which they themselves would take these actions going forward. The order in which these measures were presented was randomised between participants.

Specifically, participants were asked to both rank the actions from most effective (1) to least effective (14) at reducing personal greenhouse gas emissions, and to rate each behaviour as having either a small, moderate, or large effect on emission reductions. We then used the CO$_2$ reduction potential from Ivanova et al. \cite{ivanova_quantifying_2020}, which has been widely used for this purpose (e.g., \cite{remshard_understanding_2025, goldwert_climate_2025,cologna_knowledge_2022}), to calculate the mean absolute deviation between participants' impact \textit{rankings} and the true impact rankings (\textit{M} = 4.17, \textit{SD} = 1.11). Similarly, participants' impact \textit{ratings} were classed as either correct (0) or incorrect (1). Summing across all behaviours produced an impact assessment score ranging from 0 (i.e., the impacts of all actions were rated correctly) to 14 (i.e., the impacts of all actions were rated incorrectly; \textit{M} = 8.87, \textit{SD} = 1.85).

Participants were also asked to report how often they intend to engage in each behaviour going forward. For actions that constitute repeated behaviours (e.g., recycling, eating a vegetarian diet), the response options were presented on a Likert scale from 1 = ``Never'' to 5 = ``Always'', plus a ``Not Applicable'' option (\textit{M} = 3.84, \textit{SD} = 0.58). For one-off actions, the response options ranged from 1 = ``Very unlikely'' to 5 = ``Very likely'' (\textit{M} = 3.53, \textit{SD} = 1.07), as well as ``I already do this and will continue to do so'' and ``Not Applicable''. Additionally, participants rated how likely they would be to take three collective actions to help address climate change – contacting elected officials, donating money, or volunteering time – on a five-point Likert scale (\textit{M} = 3.98, \textit{SD} = 0.95). They were also given the opportunity to list any other actions they intend to take to address climate change in a text box. Whether participants first assessed the impacts of climate actions or reported their own behavioural engagement was randomised. 

Subsequently, we used measures from Vlasceanu et al. \cite{vlasceanu_addressing_2024} to assess participants' support for climate policies on a 100-point slider (\textit{M} = 85.32, \textit{SD} = 13.73), as well as their political orientation for social issues (\textit{M} = 39.69, \textit{SD} = 38.05) and for economic issues (\textit{M} = 42.75, \textit{SD} = 36.75). Participants were also invited to rate on a six-point scale how often they use AI tools (from ``Never'' to ``Multiple times a day''; \textit{M} = 4.10, \textit{SD} = 1.50) and how knowledgeable they are about AI (from ``I have no knowledge of AI'' to ``I am an expert in AI''; \textit{M} = 4.22, \textit{SD} = 0.91). Finally, while all participants were asked if they experienced any technical issues or had any additional feedback, participants in the personalised LLM or unspecialised LLM conditions also rated their interaction with the chatbot from Terrible (0) to Perfect (100; \textit{M} = 84.38, \textit{SD} = 18.08) and indicated their agreement with the statements ``The conversation with the chatbot was helpful'' (\textit{M} = 86.39, \textit{SD} = 18.49), ``The information provided by the chatbot was personally relevant to me'' (\textit{M} = 85.44, \textit{SD} = 18.65), and ``I felt the chatbot was trying to persuade me'' (\textit{M} = 44.52, \textit{SD} = 33.28) on 100-point slider scales.

\section*{Code and Data Availability}
All code and anonymised data needed to reproduce the analyses in this paper are available on \href{https://osf.io/9yda3}{OSF}. 

\section*{Acknowledgements}
M.R. receives support from the Yale Program on Climate Change Communication, the German Academic Scholarship Foundation, the Kurt Hahn Trust, and the NATO Science for Peace and Security Programme award number G5988. Y.K. is supported by Gates Cambridge Trust (grant OPP1144 from the Bill \& Melinda Gates Foundation) and the Alan Turing Institute's Enrichment Scheme. E.S and J.R. received support from the NATO Science for Peace and Security Programme award number G5988. The views expressed in this manuscript are solely those of the authors.

The authors would like to thank Tiancheng Hu for his thoughtful comments and suggestions. 

\section*{Author Contributions}
M.R., Y.K., E.S., and J.R. conceptualised the study. M.H.G., A.L., S.v.d.L., E.S., and J.R. provided feedback and supervision. M.R. and Y.K. analysed the data. M.R., Y.K., and J.R. led the write-up. All authors contributed to reviewing and editing the final manuscript.

\section*{Competing Interests}
The authors declare no competing interests.

\bibliographystyle{naturemag}  
\bibliography{references,yara_zotero}  


\end{document}